# Investigation of fermionic pairing on tight binding lattice for low dimensional systems

**Soumi Roy Chowdhury and Ranjan Chaudhury**

Department of Condensed Matter Physics and Material Sciences

S N Bose National Centre for Basic Sciences,

Saltlake, Sector-III, Block- JD, Kolkata-700098, India

E-mail: soumi@bose.res.in

ranjan@bose.res.in

**Abstract.** Cooper's original one pair problem in continuum is revisited here corresponding to a lattice of tight binding nature, with an aim to investigate superconductivity in low dimensional systems. An electronic type of boson mediated attraction is considered for the pairing mechanism with the non trivial energy dependence of the electronic density of states taken into account in the calculation in a rigorous way. Some of the very important electronic and optical properties of a class of quasi one dimensional organic conductors are used for the development of the formalism and calculations. The results of our calculations show that fermionic pair formation is indeed possible with some constraints. Similarities emerge in the physical properties of the electron pair formed from Cooper's treatment and ours excepting the striking difference appearing in the form of occurrences of a maximum allowed band filling for pairing and of an upper bound of the pairing energy found in our approach.

## 1. Introduction:

Superconductivity in low dimensional systems is an area of major research interest in condensed matter physics [1]. In these low dimensional superconductors, superconductivity predominantly arises as a combination of two distinct processes: - viz Cooper pair formation in a chain/ layer accompanied by inter-chain/ inter layer pair tunneling [2]. In the case of quasi-1D organic superconductors, many important questions came up regarding the pairing mechanism and the nature of the pairing symmetry. There are strong arguments both in favour of the existence of singlet pairing and of triplet pairing. We have considered electronic excitation mediated mechanism to examine pairing between two electrons in a band, involving singlets with the possible mixture of s-wave like and d-wave like character of the pair wave function. This is very much plausible in view of the experimental results on real materials [1,3-5].

**2. Mathematical formulation:**

As is well known, superconductivity doesn't occur in strictly 1D (according to Mermin Wagner theorem); however pairing between two electrons may still take place in 1D [6]. In the original Cooper's method for one pair problem in continuum, the pairing equation is

$$-\left(\frac{\hbar^2}{2m}\right)(\nabla_1^2+\nabla_2^2)\Phi(\vec{r}_1-\vec{r}_2)+V(\vec{r}_1,\vec{r}_2)\Phi(\vec{r}_1-\vec{r}_2)=E\Phi(\vec{r}_1-\vec{r}_2) \qquad (1)$$

where, $\Phi(\vec{r}_1-\vec{r}_2)$ is the spin singlet pair wave function in relative co ordinate space given by,

$\Phi(\vec{r}_1-\vec{r}_2)=\sum_{\vec{k}} a_{\vec{k}}\, e^{i\vec{k}.\vec{r}_1} e^{-i\vec{k}.\vec{r}_2}$ [Corresponding to zero momentum pairs] with the symbols having their usual meaning. In our present case of pairing corresponding to the 1D nearest neighbor tight binding lattice system, equation (1) takes the form:-

$$2a_{\vec{k}}\epsilon_k+\sum_{\vec{k}'} a_{\vec{k}'}V_{\vec{k},\vec{k}'}=Ea_{\vec{k}} \qquad (2)$$

leading to

$$2\epsilon_k a_{\vec{k}}-\sum_{\vec{k}'} a_{\vec{k}'}\frac{u}{L}=Ea_{\vec{k}} \qquad (3)$$

where

$$\epsilon_k=\epsilon_0-2t\text{Cos}(ka) \qquad (4)$$

is the single particle band energy with $'a'$ as the lattice constant. Here E is the two particle energy eigen value as before and $V_{\vec{k},\vec{k}'}$ is the Fourier transform of the attractive contact interaction $(-u\delta(\vec{r}_1-\vec{r}_2))$ being equal to $(-u/L)$ only within the small region beyond the Fermi points where the pairing would take place (usual Cooper's model); L is the size of the 1D system in consideration and $L\rightarrow\infty$ for a macroscopic system. Summing over $\vec{k}$ on both sides of equation (3) and taking the continuum limit, we get

$$1=\left(\frac{u}{L}\right)\left(\frac{L}{2\pi}\right)\int_{(\tilde{\epsilon}_{\vec{k}\,low})}^{(\tilde{\epsilon}_{\vec{k}\,up})}\frac{d\tilde{\epsilon}_k}{2at\sqrt{1-\left(\frac{\tilde{\epsilon}_k+\epsilon_F-\epsilon_0}{2t}\right)^2}\,(|W|+2\tilde{\epsilon}_k)} \qquad (5)$$

A new variable $\tilde{\epsilon}_k=\epsilon_k-\epsilon_F$ is introduced here within the standard form of 1D density of states (DOS). Besides $|W|$ is the binding energy of the two electrons where $-|W|=E-2\epsilon_F$ for $E<2\epsilon_F$, $\epsilon_F$ is the Fermi energy corresponding to a particular filling. Here $\tilde{\epsilon}_{\vec{k}\,up}-\tilde{\epsilon}_{\vec{k}\,low}$, the allowed energy range for pairing, is of the order of the band width itself. So the electronic DOS has been kept within the integrand [7]. This is a very important departure from the original Cooper pair problem where the calculation was done for a boson, having a very small energy range for attractive interaction above the Fermi surface. Therefore the energy variation of DOS was neglected there. Without the restriction imposed by the bandwidth the pairing would take place above the Fermi points throughout the energy range of $\hbar\omega_{el}$ satisfying the equation

$$\epsilon_0 - 2t\mathrm{Cos}(k_F + \Delta k)a - \{\epsilon_0 - 2t\mathrm{Cos}(k_F\, a)\} = \hbar\omega_{el} \qquad (6)$$

where $\hbar\omega_{el}$ is the characteristic energy of the exciton, mediating the pairing interaction and represents the range of the attraction above the Fermi points, $\Delta k$ is the allowed momentum space for pairing. If the boson energy, crosses the top of the electronic band, then the allowed pairing must be limited inside the available bandwidth. Thus, when $\hbar\omega_{el} > 4t(1-\delta)$, the maximum allowed energy range for the pairing of electrons is $4t(1-\delta)$ which is a measure of the available empty space in between the Fermi energy for a particular filling and the top of the band, with $\delta$ denoting the filling fraction. Therefore $(\widetilde{\epsilon}_{\vec{k}\,\mathrm{up}})$ appearing in equation (5) may be expressed as

$$(\widetilde{\epsilon}_{\vec{k}\,\mathrm{up}}) = 4t(1-\delta) + \left(\hbar\omega_{el} - 4t(1-\delta)\right)\,\theta\,(4t(1-\delta) - \hbar\omega_{el})$$

and

$$(\widetilde{\epsilon}_{\vec{k}\,\mathrm{low}}) = 0 \qquad (6.1)$$

As per chemical structure $(TMTSF)_2X$ molecule consists of two $TMTSF^+$ ions and one $X^-$ ion. It is also covalent because of large overlap, resulting from the formation of $\pi$ bonds between adjacent TMTSF molecules along the conducting axes [1]. As expected, the optical conductivity vs frequency graph shows besides the usual Drude peak, an extra hump or peak (in ev range) at position $\omega_0$, which may originate from an electronic excitation as phonons can never acquire such a high energy [8]. The non-Drude peak is often observed in the optical conductivity spectrum for many materials having similar chemical characteristics. This peak value viz. $\omega_0$ for $(TMTSF)_2X$ has been identified with $\omega_{el}$ used in our pairing calculation. Now comparing the zero and finite centre of mass momentum cases, a structured description is presented below:-

Table 1. Comparison between the zero and finite centre of mass momentum cases

| Cases | Momentum of mates | Single particle band energy | Limits of integration |
|---|---|---|---|
| Zero Centre of mass momentum | $\hbar\vec{k}$ and $-\hbar\vec{k}$ | $\widetilde{\epsilon}_k$ | $(\widetilde{\epsilon}_{\vec{k}\,up})_{zero} = 4t(1-\delta) + (\hbar\omega_{el} - 4t(1-\delta))\theta(4t(1-\delta) - \hbar\omega_{el})$ and $(\widetilde{\epsilon}_{\vec{k}\,low})_{zero} = 0$ |
| Finite centre of mass momentum | $\hbar\left(\vec{k}+\frac{\vec{q}}{2}\right)$ and $\hbar\left(-\vec{k}+\frac{\vec{q}}{2}\right)$ | $\widetilde{\epsilon}_{\vec{k}+\frac{\vec{q}}{2}} = \epsilon_0 - 2t\mathrm{Cos}\left(\vec{k}+\frac{\vec{q}}{2}\right).\vec{a} - \epsilon_F \approx \widetilde{\epsilon}_{\vec{k}} \pm atq\,\mathrm{Sin}(ka)$ (Taking the low q limit) $\widetilde{\epsilon}_{\vec{k}+\frac{\vec{q}}{2}}$ and $\widetilde{\epsilon}_{\vec{k}-\frac{\vec{q}}{2}}$ are symmetric in $\vec{k}$ space about the minimum of the band. So the calculation will be carried out taking just one among them | $(\widetilde{\epsilon}_{\vec{k}\,up})_{fin} = 4t(1-\delta) - atq\mathrm{Sin}(ka) + (\hbar\omega_{el} - 4t(1-\delta))\,\theta(4t(1-\delta) - \hbar\omega_{el})$ and $(\widetilde{\epsilon}_{\vec{k}\,low})_{fin} = atq\,\mathrm{Sin}(ka)$ |

After performing the integration by parts on the right hand side of equation (5) we get,

$$1 = \left(\frac{u}{L}\right)\left(\frac{L}{2\pi\times 2at}\right)\left\{\left[\frac{2t\mathrm{Sin}^{-1}\left(\frac{\widetilde{\epsilon}_{\vec{k}} + \epsilon_F - \epsilon_0}{2t}\right)}{(|W| + 2\widetilde{\epsilon}_{\vec{k}})}\right]_{(\widetilde{\epsilon}_{\vec{k}\,low})}^{(\widetilde{\epsilon}_{\vec{k}\,up})} + \int_{(\widetilde{\epsilon}_{\vec{k}\,low})}^{(\widetilde{\epsilon}_{\vec{k}\,up}}\left\{\frac{2}{(|W| + 2\widetilde{\epsilon}_k)^2}\int\frac{d\widetilde{\epsilon}_k}{\sqrt{1-\left(\frac{\widetilde{\epsilon}_k + \epsilon_F - \epsilon_0}{2t}\right)^2}}\right\} d\widetilde{\epsilon}_k\right\} \tag{7}$$

Now the second part above in the right hand side of eq (7) coincides with a standard form of integration, having two solutions under two different conditions [9]:-

If $a^2 > b^2$, where $a = |W| + 4t\mathrm{Cos}k_F a$ and $b = 4t$, then

$$\int\frac{\mathrm{arcSin}x\,\mathrm{d}x}{(a+bx)^2} = \frac{\mathrm{arcSin}x}{b(a+bx)} - \frac{2}{b\sqrt{a^2-b^2}}\arctan\sqrt{\frac{(a-b)(1-x)}{(a+b)(1+x)}} \tag{8}$$

(This is referred to as the 1st integral formula throughout the rest of the paper)

and if $b^2 > a^2$ then

$$\int \frac{\mathrm{arcSin}x\, \mathrm{d}x}{(\mathrm{a}+\mathrm{b}x)^2} = \frac{\mathrm{arcSin}x}{\mathrm{b}(\mathrm{a}+\mathrm{b}x)} - \frac{1}{\mathrm{b}\sqrt{\mathrm{b}^2-\mathrm{a}^2}} \ln \frac{\sqrt{(\mathrm{a}+\mathrm{b})(1+x)}+\sqrt{(\mathrm{b}-\mathrm{a})(1-x)}}{\sqrt{(\mathrm{a}+\mathrm{b})(1+x)}-\sqrt{(\mathrm{b}-\mathrm{a})(1-x)}} \tag{9}$$

(This is referred to as the 2nd integral formula throughout the rest of the paper)

The different situations, arising from the above formulae are discussed in the table below

Table 2. Summarized result obtained using 1st (left) and 2nd (right) formula

<table>
<tr>
<td>$a^2 > b^2 \rightarrow |W|^2 + 16t^2 \mathrm{Cos}^2 k_F a + 8t|W| \mathrm{Cos} k_F a > 16t^2$</td>
<td>$b^2 > a^2 \rightarrow |W|^2 + 16t^2 \mathrm{Cos}^2 k_F a + 8t|W| \mathrm{Cos} k_F a < 16t^2$</td>
</tr>
<tr>
<td>[1]The pairing energy equation according to this criterion is<br>
$$1 = \left(\frac{u}{L}\right)\left(\frac{1}{\gamma}\right) \frac{-4t}{\sqrt{(|W| + 4t\mathrm{Cos}k_F a)^2 - (4t)^2}}$$
$$\left\{\left[\tan^{-1}\left(\frac{(|W| + 4t\mathrm{Cos}k_F a - 4t)\left(1 - \frac{\tilde{\epsilon}_k - 2t\mathrm{Cos}k_F a}{2t}\right)}{(|W| + 4t\mathrm{Cos}k_F a + 4t)\left(1 + \frac{\tilde{\epsilon}_k - 2t\mathrm{Cos}k_F a}{2t}\right)}\right)^{\frac{1}{2}}\right]^{\hbar\omega_{el}} - \left[\tan^{-1}\left(\frac{(|W| + 4t\mathrm{Cos}k_F a - 4t)\left(1 - \frac{\tilde{\epsilon}_k - 2t\mathrm{Cos}k_F a}{2t}\right)}{(|W| + 4t\mathrm{Cos}k_F a + 4t)\left(1 + \frac{\tilde{\epsilon}_k - 2t\mathrm{Cos}k_F a}{2t}\right)}\right)^{\frac{1}{2}}\right]^{0}\right\}$$
where $\gamma = \frac{2\pi}{L} 2at$<br>
(10.1)</td>
<td>[1]Pairing energy equation according to this criterion is<br>
$$1 = \left(\frac{u}{L\gamma}\right)\left(\frac{-2t}{\sqrt{(4t)^2 - (|W| + 4t\mathrm{Cos}k_F a)^2}}\right) \times \left\{\ln \frac{\left[\frac{A+B}{A-B}\right]^{\tilde{\epsilon}_{\vec{k}\,up}}}{\left[\frac{A+B}{A-B}\right]^{\tilde{\epsilon}_{\vec{k}\,low}}}\right\}$$
where<br>
$$A = \left[\left(4t + |W| - 2(\epsilon_F - \epsilon_0)\right)\left(1 + \frac{\tilde{\epsilon}_k + \epsilon_F - \epsilon_0}{2t}\right)\right]^{\frac{1}{2}}$$
$$B = \left[\left(4t - |W| + 2(\epsilon_F - \epsilon_0)\right) \times \left(1 - \frac{\tilde{\epsilon}_k + \epsilon_F - \epsilon_0}{2t}\right)\right]^{\frac{1}{2}}$$
and $\gamma = \frac{2\pi}{L} 2at$ (10.2)</td>
</tr>
<tr>
<td>[2]If the filling factor tends to zero then the above inequality turns out to be $|W|^2 + 16t^2 + 8t|W| > 16t^2$. Therefore pairing easily takes place at very low filling with very small value of pairing energy. At the limit of complete filling of the band, this inequality takes the form of $|W|^2 + 16t^2 - 8t|W| > 16t^2$, implying $|W| > 8t$. This implies that the pairing energy will be very high and comparable to hopping amplitude at higher filling. Besides, the condition points towards a finite starting value of pairing energy unlike Cooper's original case.</td>
<td>[2] The above equation shows that to get the desired value of 1 at L.H.S for an infinitesimal u, the rest portion at R.H.S has to be infinite, which is achieved at $|W| \rightarrow 0$, like Cooper case. A very important consequence of our calculations from these above equations (10.1 and 10.2) is that the admissible solutions give an upper bound to |W|. This is in sharp contrast to the</td>
</tr>
</table>

| | |
|---|---|
| Thus we can't get the situation of $\lvert W\rvert^0 = 0$ here. The high filling situation is described at point 3 below. | continuum case of Cooper, where no such feature is seen. |
| [3]For having admissible values for the pairing energy equation above, only the following relations involving two sets of $\lvert W\rvert$ and $\hbar\omega_{el}$ are relevant:- <br> 1. $\begin{cases} \lvert W\rvert > 4t(1 - \mathrm{Cos}k_F a) \\ \hbar\omega_{el} < 2t(1 + \mathrm{Cos}k_F a) \end{cases}$ and 2. $\begin{cases} \lvert W\rvert > -4t(1 + \mathrm{Cos}k_F a) \\ \hbar\omega_{el} > 2t(\mathrm{Cos}k_F a - 1) \end{cases}$ <br> Besides the simultaneous non vanishing of inverse tangential parts implies <br> 3. $\lvert W\rvert + 4t\mathrm{Cos}k_F a - 4t \neq 0$ and <br> 4. $\begin{cases} 2t(1 + \mathrm{Cos}k_F a) \neq \hbar\omega_{el} \\ 2t \neq -2t\mathrm{Cos}k_F a \end{cases}$ <br> Inequality 4. is violated at $k_F a = \pi$. implying that at full filling of the band no possibility of pairing arises | [3]Same conditions or inequalities appear for having admissible values of the pairing energy equation. An important result follows from this inequality viz. $\hbar\omega_{el} < 2t(1 + \mathrm{Cos}k_F a)$ [see inequality (11) on the next page]. This implies a maximum allowed value of Fermi momentum, above which the pairing equation will be non tractable and it also relates a minimum threshold value of $'t'$ for a particular filling and bosonic energy. |

## 3. Calculation and Results

The standard functional characterization has been carried out, to extract the non-Drude peak by generating these graphs as close as possible to the experimentally obtained graph for $(TMTSF)_2X$ [7]. For this purpose the plots for 300K have been chosen because at this temperature the 1D character of the quasi 1D material is expected to be dominant.

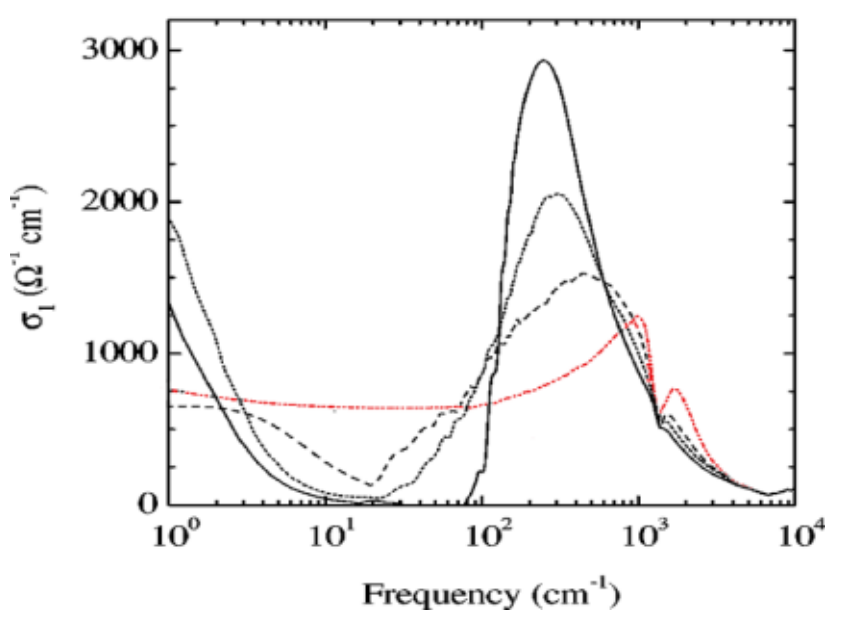

Figure 1.1. Experimentally measured absorptivity of $(TMTSF)_2ClO_4$ vs. frequency, along the highest conducting axis at different temperatures [10]

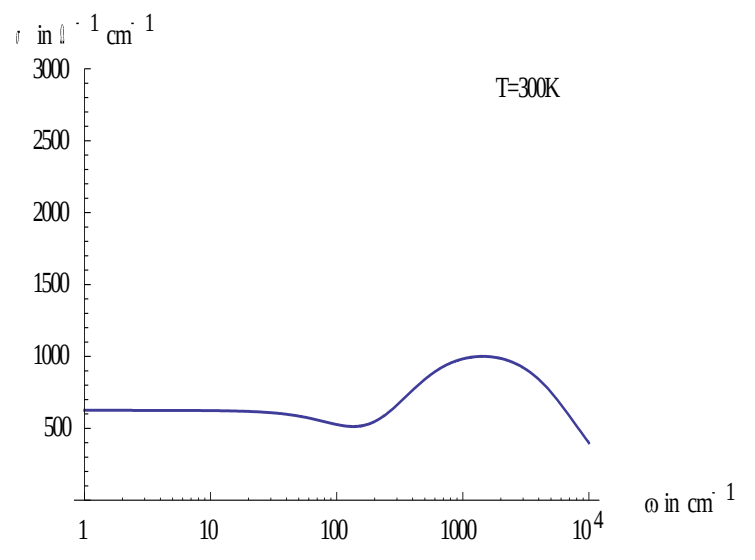

Figure 1.3.Theoretically obtained absorptivity of the same material (left) as a function of frequency at 300k

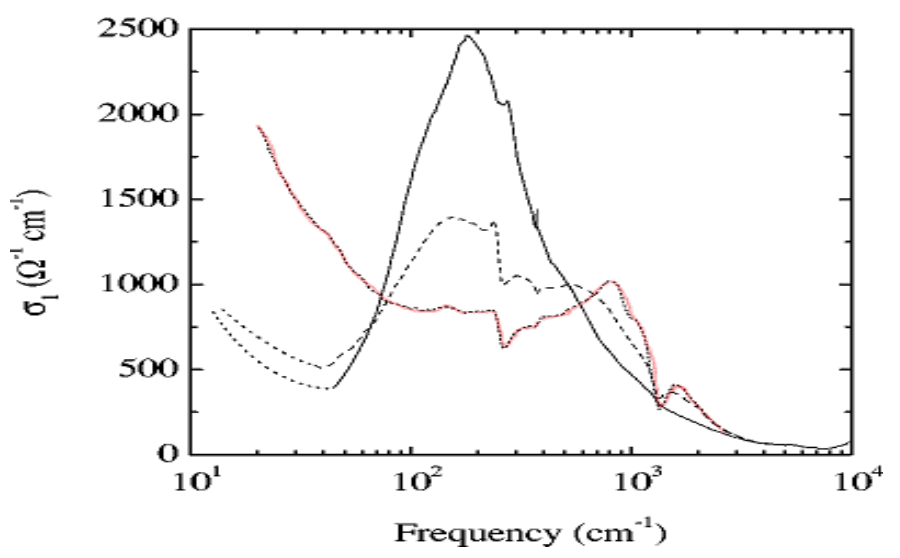

Figure 1.2. Experimentally measured absorptivity of $(TMTSF)_2AsF_6$ vs. frequency, along the highest conducting axis at different temperatures [10]

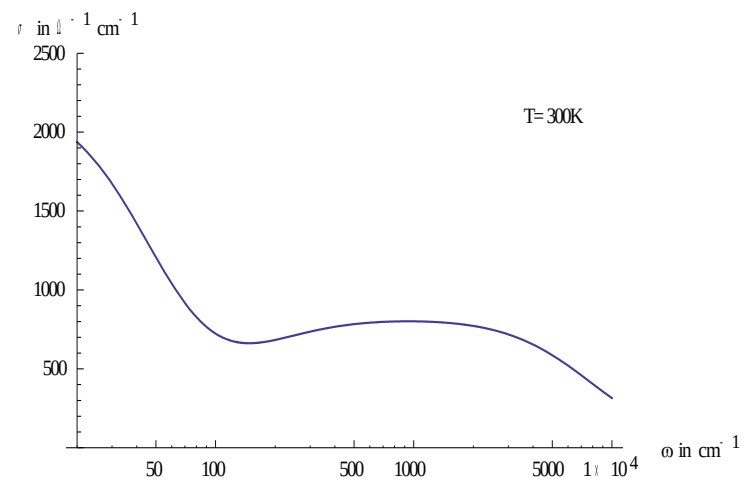

Figure1.4.Theoretically obtained absorptivity of the same material (left) as a function of frequency at 300k

Figure1. Experimentally measured and theoretically obtained absorptivity

Table3. Values of different parameters for the two following Bechhgard salts obtained by fitting

| Material | $\hbar\omega_o$ or $\hbar\omega_{el}$ |
|---|---|
| $(TMTSF)_2AsF_6$ | 1050 $cm^{-1}$ ( 0.12 ev) |
| $(TMTSF)_2ClO_4$ | 1500 $cm^{-1}$ (0.18 ev) |

The theoretically obtained optical plasma frequency ($\omega_p$) for $(TMTSF)_2AsF_6$ and $(TMTSF)_2ClO_4$ are 9022 $cm^{-1}$ and 10055 $cm^{-1}$ respectively which are close to the corresponding experimentally obtained values viz. 9900 $cm^{-1}$ and 11000 $cm^{-1}$ for these materials [10].

*3.1 Zero Centre of Mass Momentum:*

From the admissible inequalities of Table-2, we choose

$$\hbar\omega_{el} < 2t(1 + \mathrm{Cos}k_F a) \qquad (11)$$

to extract some important physics. For example if the band is half filled then in that case we must have $2t > \hbar\omega_{el}$. In fact for each filling there is a corresponding minimum allowed value of $'t'$, which is decided by the magnitude of bosonic energy. After incorporating the parameters value of ‘t’ (0.25 ev) and $\hbar\omega_{el}$ (reference Table 3) for $(TMTSF)_2ClO_4$ (as an example) the same inequality (11) leads to $k_F a < 7/9$ and hence $\delta < 7/9$ (the Fermi points being $\pm k_F$ ) [11]. Above this critical filling viz. $\delta \geq 7/9$ the pairing equation becomes non tractable. Based upon this critical value of the Fermi momentum or filling factor a distinct division of the whole band is done into two regions viz. (i) tractable regime and (ii) non-tractable regime. For the rest of the paper a dispersion around this $'t'$ value is taken to check how this affects the usual attractive interaction vs pairing energy relation and the spatial nature of coherence length. The graphs below are drawn using values of |W|, lower than its upper bound in tractable region, discussed in Table 2.

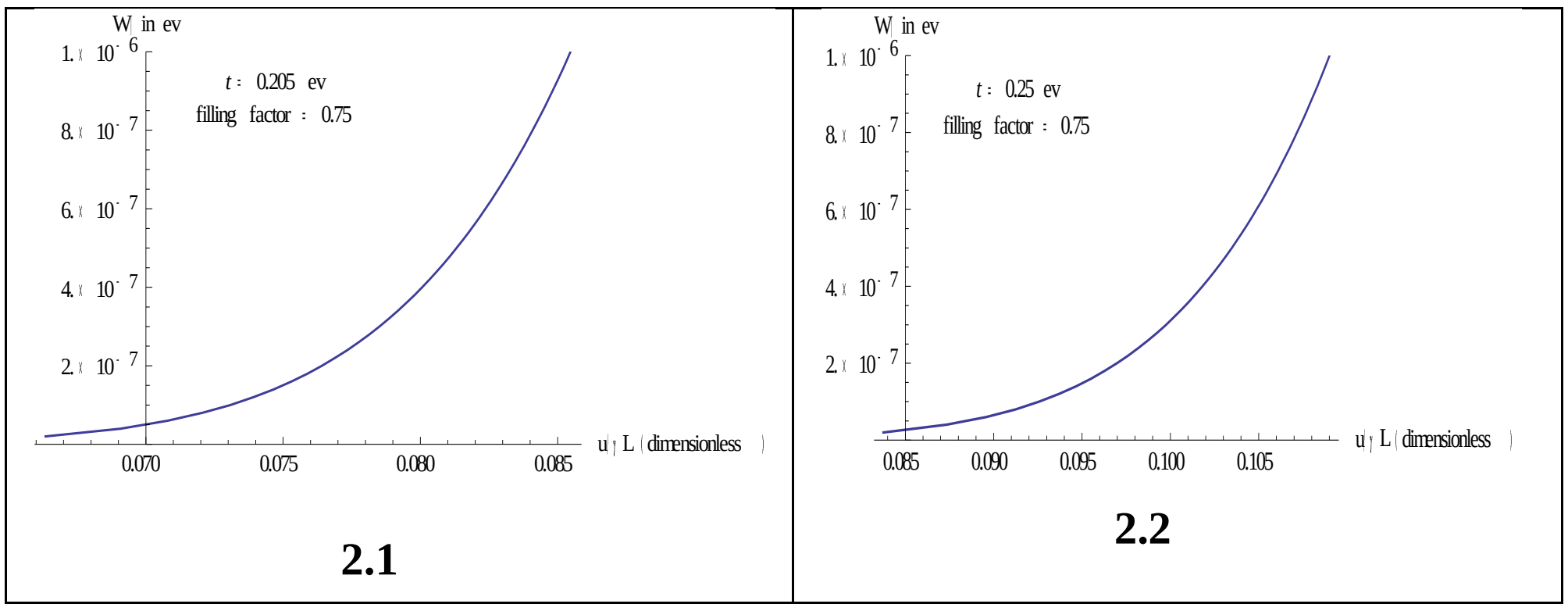

Figure 2. Plot of |W| vs. dimensionless attractive interaction for two different hopping amplitudes using 2nd integral formula.

It may be noted that although N(Є) is variable here, to have an idea about the strength of the coupling, different values of $u/\gamma L$ for a particular filling are multiplied with the DOS at the Fermi energy corresponding to that filling. The expression for the coupling constant ($\lambda$) is

$$\lambda = (u/\gamma L)N(\epsilon)$$

At a higher filling the coupling is weaker for a particular pairing energy range, as can be seen in Figures below, very similar to what happens in the original Cooper’s treatment appropriate to 1D.

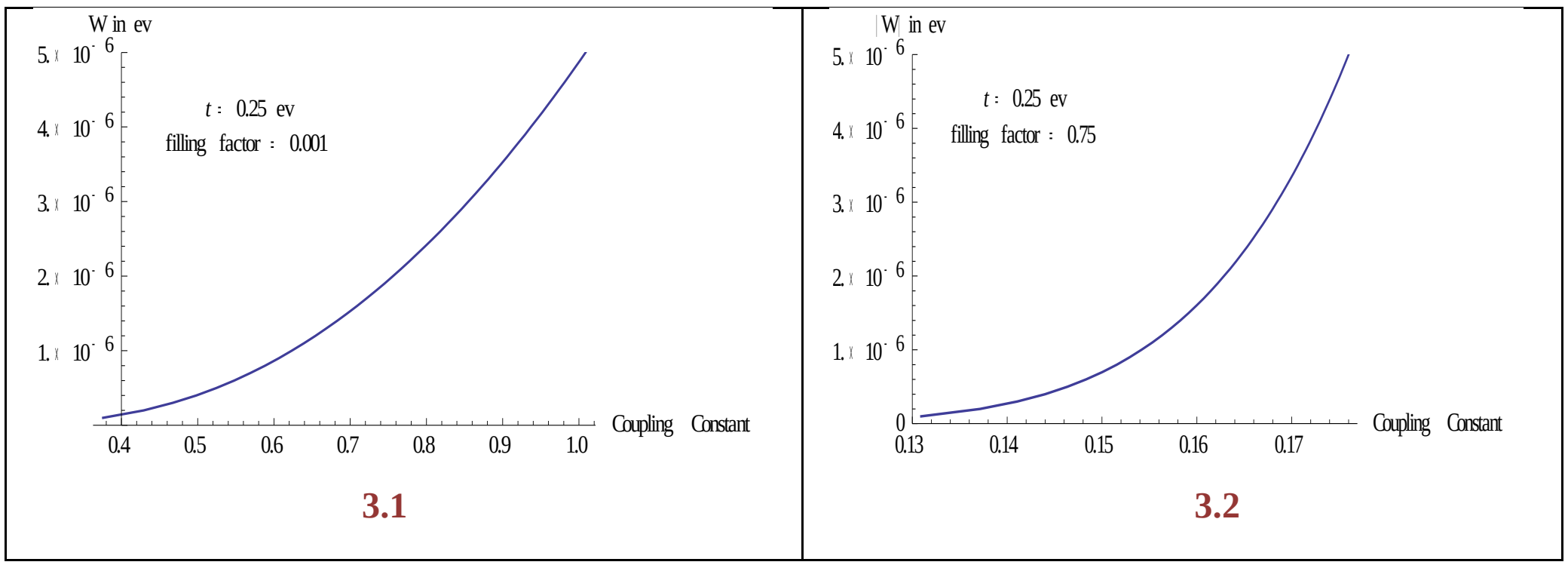

Figure 3. Graphs separately displaying variation of |W| with coupling constant for different fillings using 2nd integral formula.

*3.2 Finite centre of mass momentum*

Some mathematical results are presented below to understand the spatial nature of the pair wave function corresponding to the finite centre of mass momentum case. The maximum allowed pairing wave vector '$q_{max}$' (defined by $|W|^q = 0$ for q= $q_{max}$) gives us an estimate of the coherence length ('ξ'), which is of the order of reciprocal of $q_{max}$. We present here two tables relating ξ with different other parameters using the 2nd integral formula. We can't get an estimation of pairing involving finite centre of mass momenta by following the first integral formula as pairing energy can't be zero in this case (see point 1, Table 2). However the relation between $|W|$and q here is monotonically decreasing too.

Table 4.The calculated values of $q_{max}$ and ξ corresponding to $\delta = 0.001$ under different situations

| Value of $u/\gamma L$ (in ev) | The corresponding value of \|W\| (in ev) in case of zero centre of mass momentum | Value of 't' (in ev) | Value of $q_{max}$ (in unit of 1/a) | Value of ξ (in unit of 'a') |
|---|---|---|---|---|
| 0.00118088 | $1 \times 10^{-7}$ | 0.25 | $3.58 \times 10^{-7}$ | $0.28 \times 10^{7}$ |

Table 5.The calculated values of $q_{max}$ and ξcorresponding to $\delta = 0.75$ under different situations

| Value of $u/\gamma L$ (in ev) | The corresponding value of \|W\| (in ev) in case of zero centre of mass momentum | Value of 't' (in ev) | Value of $q_{max}$ (in unit of 1/a) | Value of ξ (in unit of 'a') |
|---|---|---|---|---|
| 0.0880748 | $5.783 \times 10^{-11}$<br>$8 \times 10^{-7}$ | 0.35<br>0.205 | $2.90 \times 10^{-10}$<br>$1.5 \times 10^{-6}$ | $0.345 \times 10^{10}$<br>$0.665 \times 10^{6}$ |
| 0.169019 | 0.0001<br>0.000811 | 0.25<br>0.205 | 0.00065142<br>0.00584 | 1535.626536<br>171.2328767 |

## 4. Discussion

1) As per electronic structure calculation one of the organic conductors $(TMTSF)_2ClO_4$ has it's conduction band 3/4th or 75% filled and shows superconductivity at ambient pressure. Our model calculation for 1D confirms pairing upto the band filling factor of 7/9 (approximately 78%) with parameters appropriate to the above system. Thus our theoretical results are consistent with this material property considering pair formation only on the highest conducting axis and pair transport along the transverse directions. It must be emphasized however that Bechhgard salts are only considered as possible support for the phenomenological scenario arising from our general formalism and calculations.

2) The magnitude of $\xi$ for higher filling (reference: Table 5) indicates a tendency towards real space like pairing if $'t'$ is reduced progressively (obeying the restriction imposed by inequality 11). This is quite plausible because if the hopping amplitude is raised it weakens the binding of the pair and the mates can now be separated by a larger distance. Therefore the coherence length increases. On the other hand the increase in the magnitude of the attractive interaction, causes an enhancement in the pairing energy resulting in reduction in the coherence length.

3) Since $\xi$ calculated here is an intra-chain parameter, the corresponding three dimensional result is difficult to be obtained directly. For that both intra chain pairing and inter chain pair hopping processes will have to be considered explicitly, so that the anisotropy of the system is taken care of in describing the process leading to superconductivity.

4)The cut offs deciding the frequency regime in the modeling for attractive pairing interaction can be determined more accurately by extracting and modeling the longitudinal dielectric function inverse $\epsilon^{-1}(q,\omega)$ from the experimental optical conductivity vs. frequency graph [12]. However a small variation in the value of bosonic energy doesn't lead to any qualitative change in our main result.

5) In our formalism the existence of forbidden bands (both above and below the band under consideration) plays a very important role in deciding the allowed pairing. We have however confined our analysis only to the particle particle pairing channel in a single electronic band. The possibility of existence of fermionic pair states or fermionic truly bound states within the forbidden bands is itself a very challenging problem and needs detailed investigation in future.

6) The Tomonaga-Luttinger Liquid model description of the normal phase as a non-Fermi liquid is less likely to be applicable here since superconductivity in these systems requires the presence of both intra-chain pair formation and inter-chain pair hopping process, making the systems very much quasi 1-D (rather than strictly 1D) like in its behaviour even in the normal phase [13, 14]

**5. Conclusion:**

The behaviour of electron on a lattice differs from that in continuum essentially due to the existence of energy bands. This leads to the contrasts in pairing properties in the two situations.The existence of an upper bound in pairing energy is one such instance. Furthermore, the existence of pairing solution also turns out to be band filling dependent in a crucial manner.